\documentclass[12pt]{article}
\usepackage{epsfig}
\newcommand{\be}{\begin{equation}}
\newcommand{\ee}{\end{equation}}
\newcommand{\bea}{\begin{eqnarray}}
\newcommand{\eea}{\end{eqnarray}}

\newcommand{\nnu}{\nonumber\\}

\newcommand{\Sigb}{{\overline\Sigma}}
\newcommand{\oot}{\overline {126}}

\begin{document}
 \vfil
 \vspace{3.5 cm} \Large{
\title{\bf  {  Pinning down the  New Minimal Supersymmetric GUT  }}
 \author{Charanjit S. Aulakh  }}
\date{}
\maketitle

\normalsize\baselineskip=15pt

 {\centerline  {\it
Dept. of Physics, Panjab University}} {\centerline{ \it
{Chandigarh, India 160014}}}

 {\centerline {\rm E-mail: aulakh@pu.ac.in }}
\vspace{1.5 cm}

\large {\centerline{\bf {ABSTRACT }}}
\normalsize\baselineskip=15pt

\vspace{1. cm} We show  that generic $ {\bf{  10\oplus 120\oplus
  {\overline {126}} }}$    fits of  fermion masses and mixings,
   using    real  superpotential couplings but
with complex `Higgs  fractions'  leading to complex yukawa
couplings in the effective MSSM,   \emph{overdetermine}(by one
extra constraint)  the  superpotential parameters
 of the new Minimal Supersymmetric SO(10) GUT\cite{nmsgut}.
 Therefore   fits should properly be done by generating
 the 24  generic fit parameters from the  23   parameters
  of the NMSGUT  superpotential, given
 $\tan\beta$ as input.  Each numerical fit      then  \emph{fully
 specifies} the parameters of the NMSGUT.   An analysis of all
 its implications, modulo only the residual uncertainty   of
 supersymmetry breaking parameters, is   now feasible.
    Thus the NMSGUT offers the possibility of a confrontation
between the  scale of gauge unification and the fit to fermion
masses due  to their extractable common dependence on the NMSGUT
parameters. If and when `smoking gun' discoveries of Supersymmetry
and Proton decay occur they will find the NMSGUT fully vulnerable
to falsification.

\normalsize\baselineskip=15pt

\newpage

\section{ Introduction}

 A series of papers over the last few
years\cite{gmblm,blmdm,core,msgreb,grimus1,nmsgut,grimus2} have
developed the   renormalizable  Supersymmetric SO(10)  GUT based
on the Higgs set ${\bf{210\oplus 10\oplus 120\oplus 126\oplus
{\overline {126}} }}$ (the so called New Minimal Supersymmetric
GUT(NMSGUT))   into a theory capable of encompassing the entire
gamut of fermion mass-mixing data in a most parameter economical
way, while preserving the traditional advantages and attractions
of renormalizable Supersymmetric
SO(10)GUTs\cite{aulmoh,ckn,rpar3,abmsv,ag1,bmsv,ag2}. The  NMSGUT
theory   has only 23 superpotential parameters and one gauge
  coupling among its `hard' parameters.
This is  one less parameter  than the original
MSGUT\cite{aulmoh,ckn,abmsv} in spite of the introduction of the
${\bf{120}}$ representation to save the feasibility of the fermion
fit \cite{gmblm,blmdm,core,msgreb,nmsgut,emerg}. It also has a
very characteristic pattern of fermion yukawas where the
${\bf{10,120}}$ couplings to ${\bf{16\cdot 16}}$ must necessarily
dominate\cite{blmdm,core,msgreb,grimus1,grimus2,nmsgut,emerg}
those of the ${\bf{{\overline {126}}}}$ to permit a   (Type I)
seesaw mechanism to generate the observed neutrino masses. This
domination  also results in  right handed neutrinos that are
lighter than the GUT scale : which may be of importance for
cosmology.

The work of \cite{grimus1,grimus2} has determined  accurate
``generic'' numerical fits of the fermion mass data by following
the strategy of coupling domination mentioned above and moreover
restricting attention to the case of real superpotential couplings
i.e only spontaneous CP violation. Although prima facie
gratifying, these successes, like those of the generic fits
without the ${\bf{120}}$\cite{allferm}, survive only with
angst\cite{grimus1,grimus2} concerning their realization in  the
full NMSGUT. In this paper we show that this angst is fully
 justified.  The generic parametrization assumes a freedom that it
 is not generically entitled to   because the underlying structure
 of the NMSGUT in fact imposes one constraint  among  the generic
 parameters which they will not, in general, satisfy. One finds that
  11 of the generic parameters\cite{grimus1}  may be expressed in terms of just
 10 NMSGUT dimensionless parameters. Therefore the numerical
 fitting program must be carried out  while respecting  one constraint.
  The only feasible way of doing this is to take the
   NMSGUT parameters as the free fitting variables instead of the generic ones.
   Thus each fit will lead to NMSGUT parameters
 sets that fully specify the theory modulo supersymmetry breaking
 uncertainties.  Indeed, in spite of the incomplete Lepton data, the
 fitting exercise will still yield complete and prima facie
 consistent parameter sets which can then be processed to make the
 latent internal contradictions, e.g. those between the
 fermion data fit and the gauge RG flow, or those
 arising   due to the use of the still incomplete fermion data,
 emerge. Thus even before supersymmetry or proton decay are
 observed we may obtain significant insights into the entire
 structure of the NMSGUT ! In this letter we give only the analytic
details of the above   arguments and leave the very involved
numerical implementation to succeeding works\cite{agrg2cum}.

\section{Basics  of NMSGUT}
 The    NMSGUT \cite{nmsgut}  is  a
  renormalizable  globally supersymmetric $SO(10)$ GUT
 whose Higgs chiral supermultiplets  consist of AM(Adjoint Multiplet) type   totally
 antisymmetric tensors : $
{\bf{210}}(\Phi_{ijkl})$,   $
{\bf{\overline{126}}}({\bf{\Sigb}}_{ijklm}),$
 ${\bf{126}} ({\bf\Sigma}_{ijklm})(i,j=1...10)$ which   break the GUT symmetry
 to the MSSM, together with Fermion mass (FM)
 Higgs {\bf{10}} (${\bf{H}}_i$) and ${\bf{120}}$($O_{ijk}$).
  The  ${\bf{\overline{126}}}$ plays a dual or AM-FM
role since  it also enables the generation of realistic charged
fermion   and    neutrino masses and mixings (via the Type I
and/or Type II Seesaw mechanisms);  three  {\bf{16}}-plets
${\bf{\Psi}_A}(A=1,2,3)$  contain the matter  including the three
conjugate neutrinos (${\bar\nu_L^A}$).
 The   superpotential   (see\cite{abmsv,ag1,bmsv,ag2,nmsgut} for
 comprehensive details ) contains the  mass parameters
 \bea
 m: {\bf{210}}^{\bf{2}} \quad ;\quad  M : {\bf{126\cdot{\overline {126}}}}
 ;\qquad M_H : {\bf{10}}^{\bf{2}};\qquad m_O :{\bf{120}}^{\bf{2}}
\eea

and trilinear couplings
  \bea
 \lambda : {\bf{210}}^{\bf{3}} \qquad ; \qquad  \eta  &:&
 {\bf{210\cdot 126\cdot{\overline {126}}}}
 ;\qquad  \gamma \oplus {\bar\gamma}  : {\bf{10 \cdot 210}\cdot(126 \oplus
{\overline {126}}})\nnu k &:& {\bf{ 10\cdot 120\cdot{ {210}}}}
\qquad;\qquad \rho :{\bf{120\cdot 120 \cdot{  { 210}}}} \nnu \zeta
&:& {\bf{120\cdot 210\cdot{  {126}}}} \qquad;\qquad\bar\zeta :
{\bf{120\cdot 210\cdot{ \overline {126}}}}
  \eea

In addition   one has two   symmetric matrices $h_{AB},f_{AB}$ of
Yukawa couplings of the the $\mathbf{10,\oot}$ Higgs multiplets to
the $\mathbf{16 .16} $ matter bilinears and one antisymmetric
matrix $g_{AB}$ for the coupling of the ${\bf{120}}$ to
$\mathbf{16 .16} $  . It was shown\cite{blmdm,core,nmsgut,grimus1}
that with only spontaneous CP violation, i.e with all the
superpotential parameters real, it is is till possible to achieve
an accurate fit of all the fermion mass data which   furthermore
 evades the difficulties
 encountered in accommodation  with the high scale structure of the MSGUT\cite{blmdm}
  provided\cite{core,msgreb,nmsgut,grimus1,grimus2,emerg}
one takes the $\mathbf{{10,120}}$ yukawa couplings to be much
larger than those of the $\mathbf{\oot}$ so that Type I neutrino
masses are enhanced.Hoever in this letter we point out that the
numerical nonlinear  fitting  procedure\cite{grimus1} must respect
a further constraint so that the use of NMSGUT parameters  becomes
mandatory.

The GUT scale vevs and therefore the mass spectrum are all
expressible in terms of a single complex parameter $x$ which is a
solution of the cubic equation

\be 8 x^3 - 15 x^2 + 14 x -3 = -\xi (1-x)^2 \label{cubic} \ee
where  $\xi ={{ \lambda M}\over {\eta m}} $.

Spontaneous CP violation implies that $x$ must lie\cite{nmsgut} on
one of the two complex solution branches
$x_\pm(\xi),(\xi\in(-27.917,\infty))$. Since $\lambda,\eta$ are
already counted as independent $x_+(\xi)$ counts for $M/m$.

\section{NMSGUT Constraints on the Grimus K$\ddot{u}$hb$\ddot{o}$ck generic
 parametrization}

 The generic parametrization\cite{grimus1} of fermion masses,
 in terms of the Yukawa-vev products   and dimensionless parameters
 arising from doublet vev ratios and vev phases,  can be  translated
 in terms of  the ``Higgs fractions''($\alpha_i,\bar\alpha_i$) determined by the fine
 tuning that keeps the MSSM pair of Higgs doublets
 light\cite{abmsv,ag1,bmsv,ag2,nmsgut}. We show that    12
 dimensionless parameters of the generic fit($\xi_{u,d,l,D},\zeta_{u,d},
r_{F,H,u,l,D,R } $,) are determined in terms of only 11
superpotential couplings ($ m_O/m,M/m,m/v,
\eta,\lambda,\zeta,\bar\zeta,\rho,\gamma,\bar\gamma,k$)of the
NMSGUT ($M_H$ is fixed by finetuning).
  Note that  even the GUT
 scale parameter(m) that sets the mass scale of all superheavy
 particles\cite{abmsv,ag1,ag2,bmsv,nmsgut} is determined by the Type I seesaw fit.  The freedom to choose  12 real
 matter fermion Yukawas are of course common to both.

 To see how these relations arise we need only compare the
 fermion mass formulae of the generic parametrization\cite{grimus1}
 with those given by us earlier for the MSGUT\cite{ag2} and the
 NMSGUT\cite{blmdm,nmsgut}. The generic parametrization reads

\begin{eqnarray}
M_d    & = & H' + e^{i\xi_d} G' + e^{i\zeta_d} F',
\label{Md} \\
M_u    & = & r_H H' + r_u\, e^{i\xi_u} G' + r_F e^{i\zeta_u} F',
\label{Mu} \\
M_\ell & = & H' + r_\ell\, e^{i\xi_\ell} G' - 3\, e^{i\zeta_d} F',
\label{Ml} \\
M_D    & = & r_H H' + r_D\, e^{i\xi_D} G' - 3\, r_F e^{i\zeta_u}
F',
\label{MD} \\
 M_{\nu}  &=& r_R ~ M_D {F'}^{-1} M_D^T. \label{Mnu}
\end{eqnarray}
The ratios $r_{H,F,D,R,u,l}$    are real by definition since they
 extracted\cite{grimus1} the phases from the VEVs. On the other hand we had
previously given   explicit formulae for all light fermion masses
in terms of the fundamental (N)MSGUT
parameters\cite{ag2,blmdm,nmsgut}.  These formulae are expressed
in terms of the so called Higgs fractions $ \alpha_i,\bar\alpha_i
;i=1..6$ which  specify\cite{nmsgut} the  MSSM  Higgs multiplet
pair $H=H^{(1)},\bar H=\bar H^{(1)}$ as a linear combination of
the 6 pairs of multiplets $h_i[1,2, 1],{\bar h}_i[1,2, -1] $
present among  the GUT Higgs fields  : \bea H=
\sum_{i=1}^{i=6}\alpha_i^* h_i \quad;\quad  \bar H=
\sum_{i=1}^{i=6}\bar\alpha_i^* \bar h_i \eea The overall phase of
the $\alpha_i,\bar\alpha_i$ is arbitrary so that we can fix
$\alpha_1,\bar\alpha_1$ to be real by a choice of this phase to
accord with the phase choice in the coefficients of
\cite{grimus1}. This modifies the normalization given
in\cite{nmsgut}.
 Explicit formulae for the Higgs fractions $\hat\alpha_i,{\hat{\bar\alpha}}_i $ in
terms of the 9 dimensionless couplings  $\tilde
m_O,\eta,\lambda,\zeta,\bar\zeta,\rho,\gamma,\bar\gamma,k$ and the
parameter $x$ which specifies the GUT scale symmetry breaking
$Susy\times  SO(10)\rightarrow MSSM $ are given in gory detail in
Appendix C of  \cite{nmsgut}. Just to illustrate the the
complexity of the formulae we give the explicit form of   the
\emph{simplest} of the un-normalized (i.e before imposing the
unitarity constraint  $\sum |\alpha_i|^2=\sum
|N\hat\alpha_i|^2=1=\sum |{\bar N}\hat{\bar{\alpha}}_i|^2$)
coefficients $\hat\alpha_1=\hat{\bar\alpha}_1$
 :
\begin{eqnarray*}
\hat{\alpha_1}& =& \hat{\bar{\alpha}}_1= ({\tilde{m_o}}^2\,{\eta
}^2\,\lambda \,{P_0}+ {\bar{\zeta}}^2\,{\zeta }^2\,\lambda \,{P_1}
+
  \tilde{m_o}\,\bar{\zeta}\,\zeta \,\eta \,\lambda \,{P_2} +
    \bar{\zeta}\,\zeta \,\eta \,\lambda \,\rho \,{P_3} +
 \tilde{ m_o}\,{\eta }^2\,\lambda \,\rho \,{P_4}
 + {\eta }^2\,\lambda
   \,{\rho }^2\,{P_5})\nnu
 P_0 &=&- 12\,{p_3}\,{p_5} t_{(1,1)}^4
 \hspace{10mm}P_1= 24\,x^3\,t_{(1,1)}\,{t_{(10,1)}}
\hspace{18mm}P_2=-24\,x \,{t_{(10,2)}}\,t_{(1,1)}^2
\\
P_3&=&4\,x \,{t_{(11,1)}}\,t_{(1,1)}^2
\hspace{10mm}P_4=8\,{p_3}\,{p_5}\,{t_{(2,3)}}\,{t_{(1,1)}^3}
\hspace{17mm}P_5=4 x^2\,{p_3}\,{p_5}\,t_{(1,1)}^4
 \end{eqnarray*}

where $t_{(m,n)}$ are polynomials in $x$ of  degree  up to 12. The
other $\alpha_i,\bar\alpha_i$ are even more complicated !

Then it  is straightforward to verify that the equations
relating\cite{nmsgut} the two parametrizations can be cast in the
form ($A,B=1,2,3$ are generation indices) : \bea H'_{AB}
&=&=2{\sqrt 2} v \bar\alpha_1\cos\beta h_{AB} \nnu F'_{AB}
&=&=4{\sqrt {\frac{2}{3}} v \bar\alpha_2\cos\beta f_{AB} }\nnu
G'_{AB} &=&=2{\sqrt {\frac{2}{3}} v |  i {\sqrt 3}\bar\alpha_5 +
\bar\alpha_6| \cos\beta g_{AB}} \label{grimyuks} \eea

for the  flavour mass matrices,

\bea  \zeta_u &=& Arg[\alpha_2]  -{\frac{\pi}{2}}\quad ;\quad
\zeta_d   =  Arg[{ {\bar\alpha_2} }] -{\frac{\pi}{2}}  \nnu
   \xi_u &=& Arg[-\sqrt{3} \alpha_5+i\alpha_6 ]
 \quad ;\quad \xi_d=Arg[{ { -\sqrt{3}\bar\alpha_5}}   +  i\bar\alpha_6  ]\nnu
  \xi_l  &=&  Arg[{ { -\sqrt{3}\bar\alpha_5 -3 i\bar\alpha_6
   }} ]
   \quad ;\quad \xi_D   =Arg[{ {-\sqrt{3} \alpha_5-3i  \alpha_6
   }} ]\label{phases} \eea

for the phases, and

\bea r_H &=&   {\frac{\alpha_1}{\bar\alpha_1}} \tan \beta \quad ;
\quad r_F  =    |{\frac{\alpha_2}{\bar\alpha_2}}| \tan \beta  \nnu
r_u &=& |{\frac{   \sqrt{3} \alpha_5  -i\alpha_6   } {
\sqrt{3}\bar\alpha_5 -i\bar\alpha_6
    }} | \tan\beta\nnu
 r_l &=& |{\frac{ \sqrt{3}\bar\alpha_5 + 3 i\bar\alpha_6  } { \sqrt{3}\bar\alpha_5 - i\bar\alpha_6
    }} | \tan\beta\nnu
 r_D &=& |{\frac{  \sqrt{3} \alpha_5  +3 i  \alpha_6  }
     { \sqrt{3}\bar\alpha_5  -i  \bar\alpha_6  }}| \tan\beta \nnu
     r_R &=&  {{\frac{|\bar\alpha_2|\lambda \cos\beta}{{2\sqrt 3}
     |\tilde{\bar\sigma}|}  }}
     {\frac{v}{m}}  \label{vevrats} \eea
for the ratios of vevs in the parametrization of \cite{grimus1}.
Here $\tilde{\bar\sigma}$ is the dimensionless ${\bf\oot}$ vev in
units of $m/\lambda$. Note that in
eqns(\ref{grimyuks},\ref{phases},\ref{vevrats}) the values of
$v,\tan\beta$ (or equivalently $v_u=v \sin\beta,v_d=v\cos\beta $)
are the renormalized values   at the GUT scale obtained from  RG
flow. To two loops the RG equations\cite{martinRG} for the
couplings and vevs  depend only on the gauge and matter Yukawa
couplings together with the input initial value of  $ \tan\beta$.
This is what insulates the deductions from the superpotential
parameter set from the depredations of the chaos inducing
ignorance regarding supersymmetry breaking parameters, making
possible the ambitions of the fitting program.

The first eleven of the
eqns(\ref{grimyuks},\ref{phases},\ref{vevrats} )   fix 11 of the
coefficients of the parametrization in \cite{grimus1} in terms of
  10 dimensionless GUT parameters \emph{and hence must obey one
constraint}.   Then it follows that a fermion fit found (e.g via
the downhill simplex method\cite{grimus1,grimus2}) by freely
varying   the 12 parameters  $\xi_{u,d,l,D},\zeta_{u,d},
r_{F,H,u,l,D,R} $    and the flavour mass  matrices $H',F',G'$(12
parameters),   \emph{must be checked after the fit is achieved to
verify whether the constraint is satisfied}. However this is
easier said than done because the
  equations in terms of the GUT parameters are are so hopelessly
nonlinear\cite{nmsgut}. Moreover it is clear that
\emph{generically} there is no reason to expect that the
constraint will be satisfied. \emph{Therefore the correct
procedure to determine   fermion fits that do  not fall foul  of
the necessity to respect the constraint imposed by the NMSGUT is
obviously to take the NMSGUT superpotential parameters as the the
freely variable ones.} This is the main conclusion of this letter.
The actual implementation of the very lengthy  numerical codes in
terms of the GUT parameters is currently being carried out and
will be reported separately\cite{agrg2cum}. The reason for the
extra computational burden is simply the highly complicated
expressions\cite{nmsgut} for the Higgs fractions
$\alpha_i,\bar\alpha_i$ in terms of the GUT parameters. In the
next section we conclude with a discussion of where we expect such
a  modified and consistent fitting program to leave us vis a vis
the NMSGUT.

\section{Discussion}

In this  letter we   pointed out the correct procedure for fitting
the only data currently available for constraining any GUT beyond
the basic requirements of gauge unification : the low energy
fermion masses and mixing. Although the NMSGUT has 23
superpotential parameters(after the fine tuning to keep  a Higgs
pair light) \emph{all} of which enter the fermion mass formulae,
the number of data values is smaller in number. In the most
favourable situation one would have knowledge of 12 fermion
masses, 4  CKM parameters, $6$ PMNS parameters i.e 22 data in all.
Unfortunately, however, of these 22 there is no prospect in sight
for measuring the two Majorana phases at present.  The leptonic
mixing angle $\theta_{13}^l$, the Dirac PMNS phase $\delta_\ell$
and the combination of the neutrino masses effective in
neutrinoless double beta decay may become available in the near
future and may, in any case, be plausibly bounded from above.
Moreover an  estimate of neutrino masses may also emerge from
cosmology. Practically speaking one must then fit to 11 fermion
masses(since only the neutrino mass squared differences are known)
and 7 mixing data. Thus  -for the present- one needs to fit 18
data values with 23 parameters through highly nonlinear relations.
The only practicable way of doing this is through a numerical
fitting procedure such as the downhill simplex
algorithm\cite{grimus1,grimus2} or possibly using some hybrid
procedure of an ansatz regarding the hierarchy structure combined
with numerical fitting\cite{core,msgreb}. The accurate generic
fits found so far\cite{grimus1,grimus2} cannot be accommodated in
the NMSGUT unless they happen to satisfy the constraint tmposed by
the NMSGUT ; but that is very difficult -and not really
worthwhile- to verify.  Instead one should perform the fitting by
using the NMSGUT parameters : which are truly independent.   Thus
what emerges -as has been argued here- is not just a set of
plausible values for the generic fitting parameters(for each
assumed value of $\tan\beta$) but rather a complete candidate
parameter set for the 23 NMSGUT parameters.

The very `ease' of the numerical solution has however a high
price, in that one has practically no understanding of the global
structure of the solution space. Moreover, since one will perform
the fit without constraining  the  parameters (such as leptonic
Majorana phases) whose value one does not know at present, it
follows that a variety of fits corresponding to different
resultant values of such parameters  may in fact be possible. In
addition the variation induced by different choices of
$\tan\beta$, as well as the RG amplified uncertainties in the
fermion data, must also be considered.

However the blindness of the fitting procedure -if successful-
opens an interesting possibility : since all GUT parameters are
determined by the fermion fit alone we may confront them with the
 accurate data on gauge couplings at low energy by supplying these
 fermion data determined values to the RG flow equations for gauge
 coupling unification. The consistency between these independent
 determinations of the unification parameters will then serve as
 an important check on the plausibility of the theory and may even
 limit the viable values of $\tan\beta$.

 The complete determination of the GUT parameters by the fermion
 fit will also make possible an explicit evaluation of the proton
 decay rate. Explicit formulae for the effective Superpotential
  controlling proton decay in terms of the NMSGUT  matter fermion
  yukawa couplings and the Higgs fractions have been derived in
   \cite{ag1,ag2,nmsgut}. Of course the rate calculation  will still be afflicted by the usual
 superpartner mass and  mixing uncertainties\cite{gorborpav} but the actual
 decay rate corresponding to various Susy breaking spectra models
 (gravity mediation, gauge mediation, Higgs mediation etc) which are
 available in the literature may be used to constrain the
 predictions.  Thus our analysis indicates that the road to a
 comprehensive numerical analysis of all aspects and predictions
 of the NMSGUT appears to be open.  If the theory proves to
 generate consistent data points in spite of the nontrivial
 consistency check between fermion data and gauge unification then
 it will be an attractive contender for confronting the smoking
 gun data of Supersymmetry and proton decay discovery, if and
 when they arrive. This scenario amounts to an unfolding of the
 intricate(``ouroborotic") interplay of the the tiny(neutrino) and
 superheavy  mass  scales which, in SO(10),  are linked
 by the seesaw  in  the very  guts of the NMSGUT.

\section{Acknowledgments}
 \vspace{ .5 true cm}
C.S.A acknowledges  encouraging correspondence with F. Vissani.
The work of C.S.A was supported by a grant No SR/S2/HEP-11/2005
 from the Department of Science and Technology of the Govt. of
 India and that of S.K.G by a University Grants Commission Junior
 Research fellowship.

\end{document}